\documentclass[reprint,amsmath,amssymb,aps]{revtex4-1}

\usepackage[utf8]{inputenc}

\usepackage{graphicx}

\usepackage{hyperref}

\usepackage{amssymb}
\usepackage{color}

\usepackage{subfigure}
\usepackage{caption}
\usepackage[english]{babel}

\hypersetup{
	colorlinks = true,
	urlcolor   = blue,
	linkcolor  = blue,
	citecolor  = blue
}
 
\begin{document}

\title{Single atom movement with dynamic holographic optical tweezers}

\author{S.R.Samoylenko}
\author{A.V.Lisitsin}
\author{D.Schepanovich}
\author{I.B.Bobrov}
\author{S.S.Straupe}
\email{straups@quantum.msu.ru}
\author{S.P.Kulik}
\affiliation{Quantum Technologies Center, Department of Physics, M.V.Lomonosov Moscow State University, Moscow, Russia}

\begin{abstract}
	We report an experimental implementation of dynamical holographic tweezers for single trapped atoms. The tweezers are realized with dynamical phase holograms displayed on the liquid crystal spatial light modulator. We experimentally demonstrate the possibility to trap and move single rubidium atoms with such dynamic potentials, and study its limitations. Our results suggest that high probability transfer of single atoms in the tweezers may be performed in large steps, much larger then the trap waist. We discuss intensity-flicker in holographic traps and techniques for its suppression. Loss and heating rates in dynamic tweezers are measured and no excess loss or heating is observed in comparison with static traps.
\end{abstract}


\maketitle

\section{Introduction}
\label{Introduction}

Cold neutral atoms in optical microtraps are a promising platform for quantum simulation and computing. Recent advances have demonstrated the utility of this platform for simulation of many-body physics \cite{Lukin_Nature2017,Browaeys_Nature2016}, as well as its promise for realization of high-fidelity gates for universal quantum computation \cite{Lukin_PRL2019,Saffman_PRL2019}. The main advantages of the microtraps platform are scalability \cite{Birkl_PRL2019}, long coherence times (up to hundreds of ms) \cite{yang2016coherence}, high-fidelity readout with low cross-talk \cite{Saffman_PRL2017} and prospects for achieving high-fidelity logical gates \cite{Saffman_JPB2016}. 

A common way to produce arrays of trapped single atoms is to use holographically generated arrays of tightly focused far-off resonance dipole traps \cite{Grangier_JOSAB2004,nogrette2014single,Nakagawa_OptExp2016,Kuhn_NJP2018}. An alternative way is based on creating tweezer arrays with acousto-optic deflectors driven by multi-frequency signals \cite{Lukin_Science2016,Regal_PRL2015,Endres_PRX2018}. The advantage of holographic arrays is the realtive easiness of crating arbitrary two-dimensional \cite{nogrette2014single} and three-dimensional \cite{barredo2018synthetic} patterns of microtraps, as well as the unique capability of designing traps of complicated shapes \cite{Fatemi_PRA2007,Hoogerland_RSI2017}. 

Single atom trapping in tightly focused dipole traps is facilitated by light-assisted collisions which for small ($\sim 1$~$\mu$m waist) traps lead to the so-called collisional blockade effect \cite{schlosser2002collisional}. In the blockade regime the probability of finding two atoms in a single trap is negligible, due to strong two-atom loss caused by inelastic collisions \cite{Grangier_Nature2001,schlosser2002collisional}. Unfortunately, the time-averaged probability of single atom trapping with a tweezer in a collisional blockade regime is only about $P=0.5$ when the loading is performed directly form the magneto-optical trap. It prevents one from creating completely regular structures filled with cold single atoms which are desirable for quantum computing and simulations. Although there are several techniques that could increase the loading probability \cite{Andersen_NaturePhys2010,Regal_PRL2015}, the demonstration of a fully deterministic loading process in large arrays is still a challenging experimental problem.   

An accepted way to resolve this problem is to create randomly filled array of tweezers, experimentally determine the tweezers that are filled with single atoms and to reconfigure these tweezers into a desired structure. Fully loaded 2D and 3D structures consisting of up to 50 atoms were created with an additional fully steerable dipole trap \cite{barredo2016atom, barredo2018synthetic} or using computer generated dynamic holographic masks displayed on a spatial light modulator (SLM) \cite{kim2016situ,lee2016three, kim2019gerchberg}. 

The main advantage of dynamic holographic masks is the possibility to move multiple atoms at once \cite{Ahn_PRA2017}, but this method has some serious limitations. The main issue is that the achievable speed of atom steering is strictly limited by the SLM refresh rate, which is usually very low for typical liquid crystal devices. At the same time, a lot of computational power is required to calculate phase masks on line with a suitable frame rate. The last but not the least drawback is the intensity flicker that arises while the image displayed by the SLM updates. The flicker arises because of the finite response time of the device and is unavoidable for all types of SLMs, however, it is most dramatically pronounced in liquid crystals based modulators. There are methods for flicker damping \cite{persson2010minimizing}, however, it is hardly possible to completely remove intensity modulation for dynamic holograms. 

In this work we estimated the influence of intensity flicker on atom loss during the process of atom movement. We measure atom survival probabilities in the dipole trap at the end of movement, realized using dynamic phase masks generated by the weighted GS-algorithm with and without phase change checking step. We show, that it is possible to achieve smooth motion with no additional loss, as compared to the atom lifetime in the static dipole trap. Remarkably, the motion may be carried on by shifting the trap in quite large steps, comparable to the trap radius. This is especially advantageous for slow devices, such as liquid crystal SLMs. We also studied the energy distribution of the atom before and after the motion, and found no evidence of additional heating in the dynamical trap.

\section{Experimental setup}
\label{Experimental setup}
\begin{figure}
	\centering
	\includegraphics[width=\linewidth]{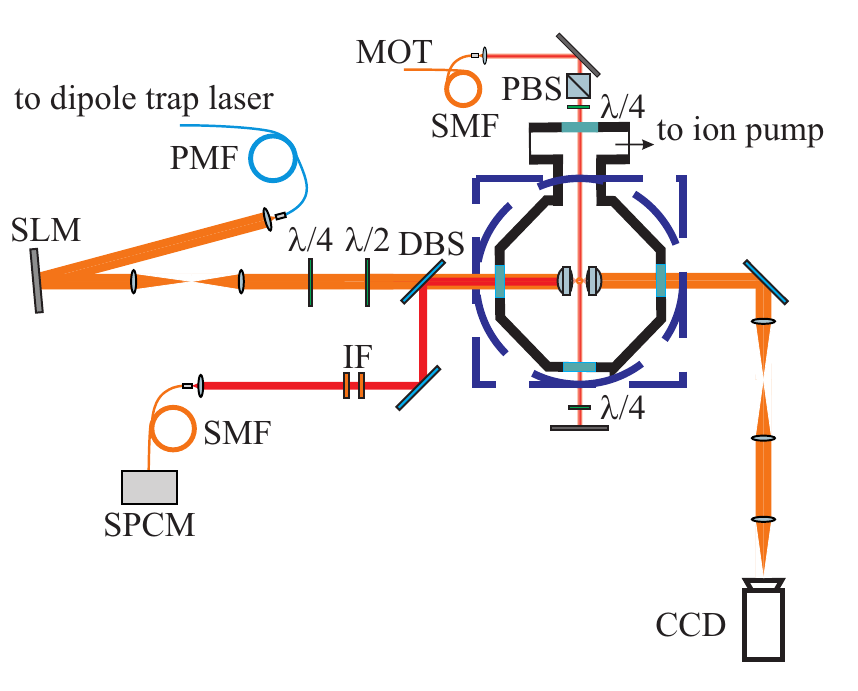}
    \caption{Experimental setup scheme. $^{87}$Rb atoms are cooled in a magneto-optical trap and single atoms are loaded in a far-off resonance optical tweezer formed by tightly focusing an 852~nm beam with an aspherical lens with $NA = 0.77$ and effective focal length $f = 3.1$ mm to a waist of 0.97 $\mu$m. Power required for a single trap is about 3.5 mW. To produce dipole trap arrays we use an SLM which is conjugated with the surface of the aspheric lens by a 0.83x telescope ($f_1$ = 300~mm, $f_2$ = 250~mm). The atom fluorescence signal is collected by the same aspheric lens, coupled to a single-mode fiber and then detected by a avalanche photodiode single photon counting module. The transmitted dipole trap beam is then collected by another aspheric lens and imaged into a CCD-camera. 
   	The optical system consisting of $f_3$, $f_4$ and $f_5$ lenses ( $f_3$ = 250 mm, $f_4$ = 400 mm, $f_5$ = 500 mm) images the dipole traps plane onto the CCD camera with a 137.5X magnification.}
    \label{fig:setup}
\end{figure}

Figure \ref{fig:setup} shows a schematic image of the experimental setup that we use to cool and trap single atoms of $^{87}$Rb in microscopic dipole traps. It consists of an ultrahigh-vacuum chamber that maintains pressure of about $10^{-10}$ mbar and contains a pair of high NA aspheric lenses to form the tightly focused dipole trap. The atoms are cooled in a MOT formed by three retroreflected laser beams, the one lying in the figure plane is shown, while the other two are in the orthogonal plane and have an angle of $\sim 20^\circ$ limited by the aperture of the lenses. The cooling beams are red-detuned by $\delta = 24$ MHz from $F=2 \rightarrow F=3$ transition of the $^{87}$Rb D2 line, while the repump beams are resonant with $F=1 \rightarrow F=2$ transition of the D1 line. 

We use a beam of a 852 nm diode laser to trap single atoms from the cold atomic cloud in a microscopic far-off-resonant trap. The beam is focused to a waist with $1/e^2$ radius of 0.97 $\mu$m by an aspheric lens with an effective focal length of $f = 3.1$ mm and $\mathrm{NA} = 0.77$. The power required to trap a single atom at this detuning is about $\approx$ 3.5~mW, which corresponds to a trap depth of $U_0 \approx 1.08$ mK. We measured the longitudinal and transverse oscillation frequencies for a trapped atom to be $w_l$ = 20.8 kHz and $w_t$ = 105.2 kHz, respectively.

We use a reflective LCoS SLM (Hamamatsu X10468-02) that is placed in the path of dipole laser beam to produce dynamical holographic optical trap arrays. An 0.83x telescope that is placed between the SLM and the vacuum chamber is used to conjugate the SLM plane and the plane of an aspherical lens. Single atoms' fluorescence signal is collected by the same lens and passes through a dichroic beamsplitter and a pair of interference filters with a bandwidth of 3~nm to cut-off the reflected radiation from the dipole trap laser. Filtered signal is then coupled to a single-mode optical fiber and detected by an avalanche photo diode single photon counting module.

\section{Phase hologram generation}
\label{GS}

\begin{figure}
	\centering
	\includegraphics[width=\linewidth]{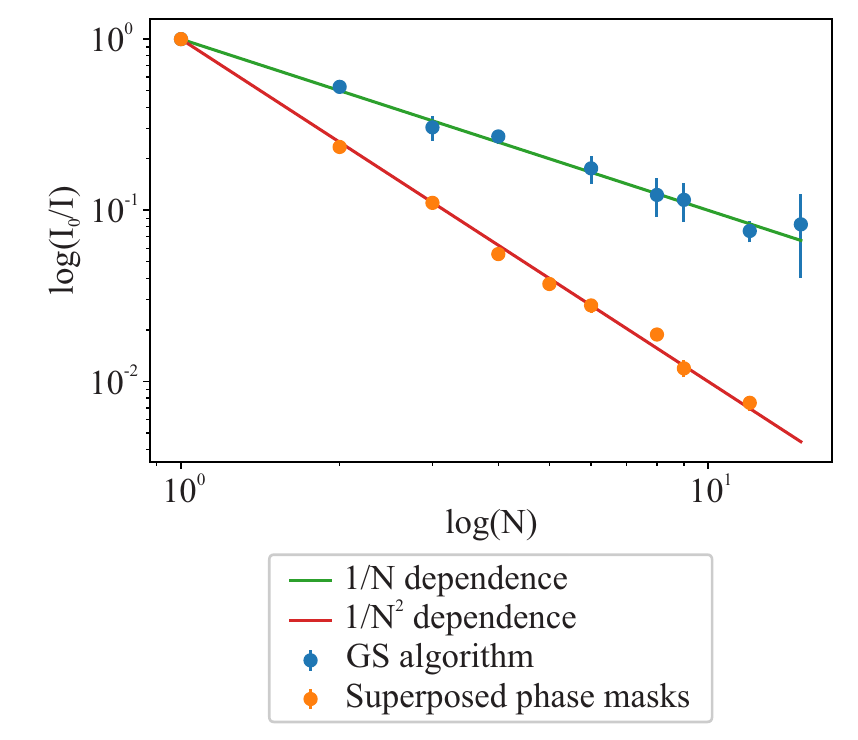}
    \caption{Scaling of the intensity per single trap in an array of $N$ traps formed by a phase hologram from a single fixed intensity laser beam. Blue dots -- experimental data for the Gerchberg-Saxton algorithm, orange dots -- experimental data for a superposed phase gratings method, green line is a $1/N$ dependence, red line -- $1/N^2$ dependence.}
    \label{fig:GS_vs_an}
\end{figure}

The Gerchberg-Saxton algorithm is one of the most widespread Fourier-transform based algorithms used to produce phase holograms for laser beam-shaping due to high achievable diffraction efficiency and simple implementation. It allows one to produce holographic arrays of traps with a single trap intensity scaling with the number of traps as $I/N$, where $I$ is the overall laser beam intensity. It's a lot more efficient in comparison to a straightforward method of superposed diffraction gratings \cite{lee2016three} , which has an $I/N^2$ intensity scaling, as shown in Figure \ref{fig:GS_vs_an}). 

\begin{figure*}
	\centering
	\includegraphics[width=\linewidth]{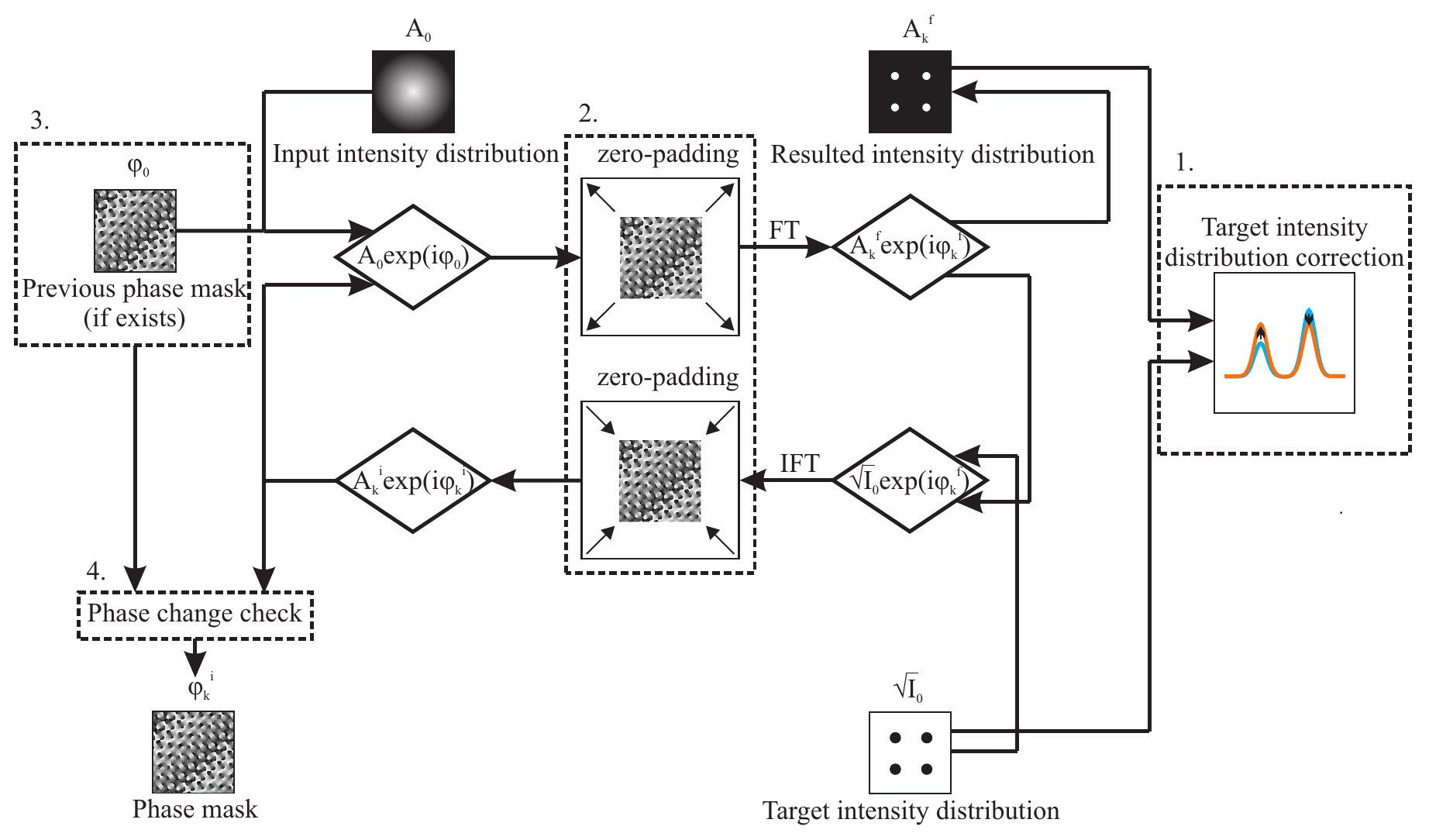}
    \caption{Modified Gerchberg-Saxton algorithm. Our modifications are highlighted by dashed boxes. 1. Weighing via correction of the target intensity distribution increased algorithm's convergence and minimized the dispersion of traps intensities. 2. Zero-padding was used to increase the resolution of the target intensity distribution. 3. The previous phase mask is used as an initial phase distribution to generate series of smoothly varying holograms for dynamic motion of the tweezers. 4. The phase change checking step. The last two steps minimize the intensity flicker during the phase masks switching.}
    \label{fig:GS_A}
\end{figure*}

Schematically the Gerchberg-Saxton algorithm is illustrated in Figure~\ref{fig:GS_A}. It consists of the following steps:

\begin{enumerate}
    \item Amplitude distribution of an incident beam  $A_0$ and an initial phase distribution of the hologram $\phi_0$ are set;
    \item The amplitude distribution if Fourier transformed $FT[A_0\exp(i\phi_0)]\rightarrow A_k^f\exp(i\phi_k^f)$. Here $A_k^f$ is expected far-field amplitude distribution. If $A_k^f$ doesn't match the target distribution $\sqrt{I_t}$ evaluation proceeds to the next step;
    \item The amplitude in the far field is replaced by the target distribution and inverse-Fourier transformed $IFT[\sqrt{I_0}\exp(i\phi_k^f)]\rightarrow A_k^i\exp(i\phi_k^i)$;
    \item The phase distribution is updated $\phi_{k-1}^i\rightarrow \phi_k^i$ and the algorithm returns to step 2;
    \item If $A^f$ matches the target intensity distribution $\sqrt{I_0}$, the loop ends. The last calculated phase distribution $\phi^i$ is sent to the SLM.
\end{enumerate}

Our modifications to the original GS algorithm are highlighted in Figure~\ref{fig:GS_A} by dashed boxes. For convenience they are listed here in the same order as they are enumerated in Figure~\ref{fig:GS_A}:
\begin{enumerate}
    \item To improve convergence of the algorithm and to minimize the deviation of the trap intensities from the mean we introduce weighing at each iteration \cite{nogrette2014single}. The essence of this method is the following: when the expected intensity distribution $A_k^f$ is calculated at step 2, the intensity of every single trap $I_i$ in the desired intensity distribution $\sqrt{I_0}$ is replaced with $\bar{I}/(1-G(1-I_i'/\bar{I}))$, where $\bar{I}$ is the calculated mean value of the trap intensities, $I_i'$ is the calculated intensity of i-th trap and $G$ is a gain coefficient which in our case was taken to be $G=0.8$.
    \item Zero-padding. It's a widespread signal processing method usually used to increase the number of samples in the frequency domain. It was shown in \cite{kim2019gerchberg} that the use of the zero-padding increases the resolution of the target intensity distribution and thus ensures that the waist of the dipole trap beam is limited only by the optics, not the algorithm.
    \item To minimize the intensity flicker of dynamic phase holograms we used a phase induction method, the main idea of which is to take the previous phase hologram as the initial phase distribution for the next one. It was shown in \cite{kim2019gerchberg} that this method significantly decreases the amount of flicker.
    \item For further flicker reduction we used a restricted phase change method \cite{persson2010minimizing}. It could be described as follows:
    \begin{enumerate}
        \item Two subsequent masks are taken;
        \item Phase change for every pixel is calculated;
        \item If the phase change $\Delta \phi$ is greater than $2\pi\alpha$, the value of the phase in the following mask is replaced by the value of phase of an appropriate pixel from the previous mask. $\alpha$ is a variable coefficient which can be tuned to take any value ranging from 0 to 1.
    \end{enumerate}
\end{enumerate}

\begin{figure}
	\centering
	\includegraphics[width=\linewidth]{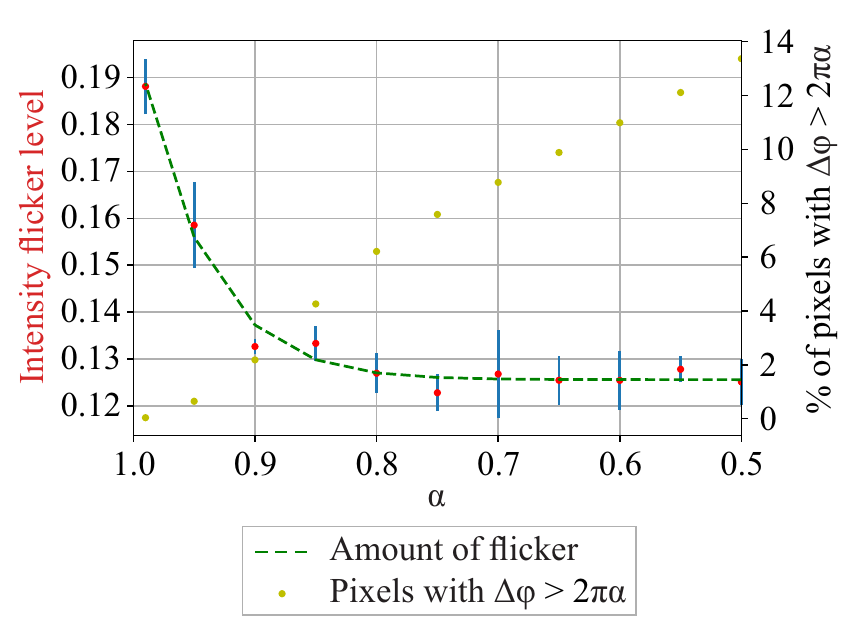}
    \caption{Results of the restricted phase change method use for hologram generation. Yellow dots -- the number of pixels in which the phase change has exceeded the level of $2\pi\alpha$, red dots -- the dependence of the flicker strength $(I_{max} - I_{min}) / I_{max}$ on the parameter $\alpha$ limiting the phase change. Green dashed line is a fit by a function $a\exp{b(\alpha - 1)} + c$ with coefficients  $a = 0.076$, $b = 18.566$, $c = 0.125$.}
    \label{fig:damping}
\end{figure}

\begin{figure}
\centering \subfigure[]{
\includegraphics[width=\linewidth]{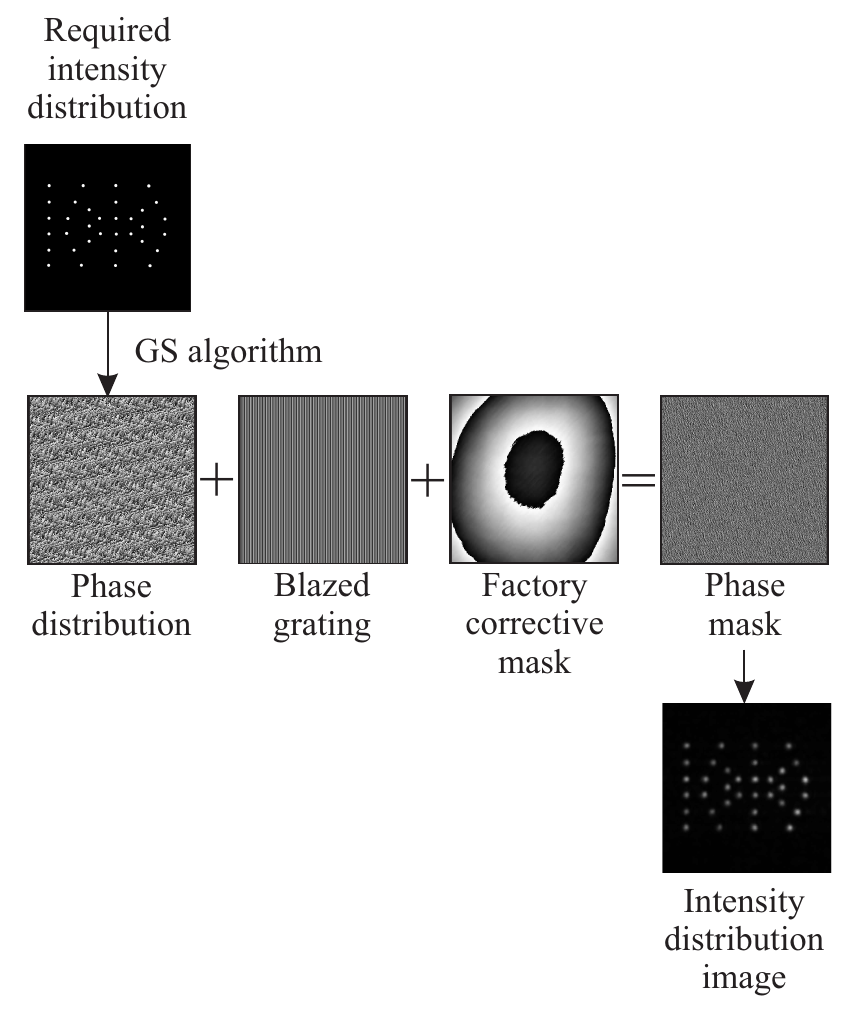} \label{fig:gs-pros} }  
\subfigure[]{
\includegraphics[width=\linewidth]{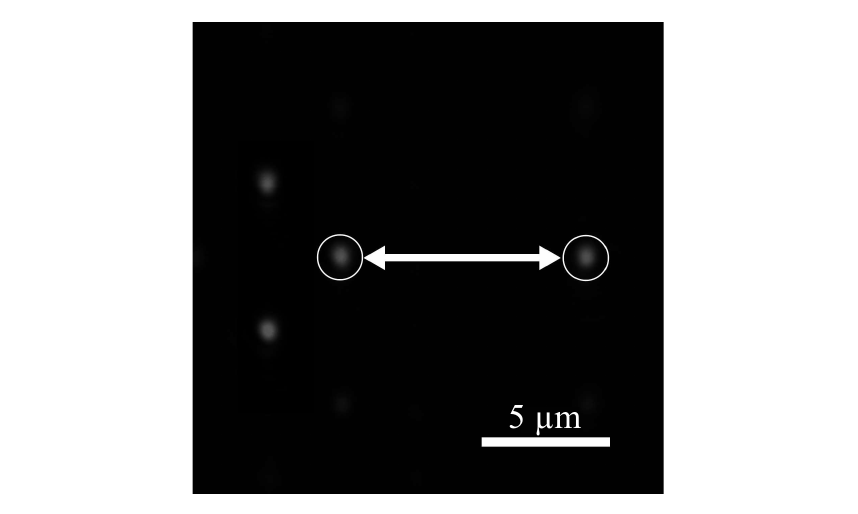} \label{fig:triangle_step} }
\caption{\subref{fig:gs-pros} The phase hologram used to create twezzer arrays in our experiments is the sum of a Gerchberg-Saxton calculated phase distribution, a blazed grating and a factory corrective mask. \subref{fig:triangle_step} The trajectory of a dipole trap that was used in atom motion experiments. The triangular array was chosen as a simplest 2D intensity distribution.} \label{fig:triangle}
\end{figure}

Dependence of the intensity flicker strength on $\alpha$ is shown in Figure~\ref{fig:damping}. It was measured for two subsequent phase masks from the sequence used later to form a triangle of dipole traps with one of the traps moving (Figure~\ref{fig:triangle_step}). The triangle shaped intensity distribution was chosen as a simplest nontrivial 2D intensity distribution. One of the traps was moving back and forth at some distance during the measurement. Dipole trap laser radiation was detected by a photodiode which was placed at the focal plane of a $f_3$ lens (Figure \ref{fig:setup}). The distance between the dipole traps in the image plane was much smaller than the size of the photodiode, allowing all to detect the total power of the diffracted light.  The flicker strength was quantified as $(I_{max} - I_{min}) / I_{max}$, where $I_{max}$ is the total intensity of the diffracted light and $I_{min}$ the total intensity of light detected during the interval between the two subsequent frames.

Finally, we add a factory preset corrective mask to the hologram calculated by the Gerchberg-Saxton algorithm to compensate for the curvature and irregularity of the modulator surface. We also add a blazed grating to our mask to eliminate the undiffracted light, see Figure~\ref{fig:gs-pros}.

\section{Experimental results}
\label{Experiment}
\begin{figure}
	\centering \subfigure[]{
		\includegraphics[width=\linewidth]{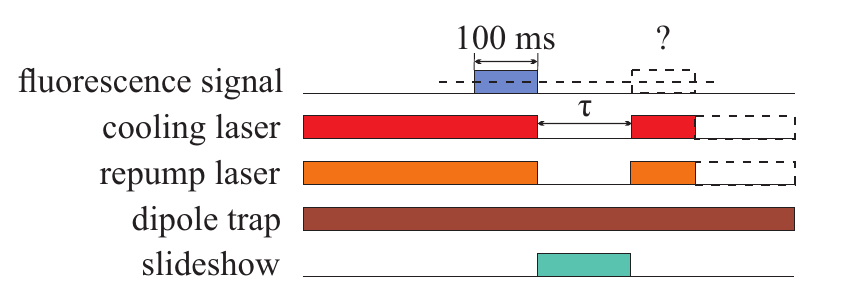} \label{fig:lifetime_sequence} }  
	\subfigure[]{
		\includegraphics[width=\linewidth]{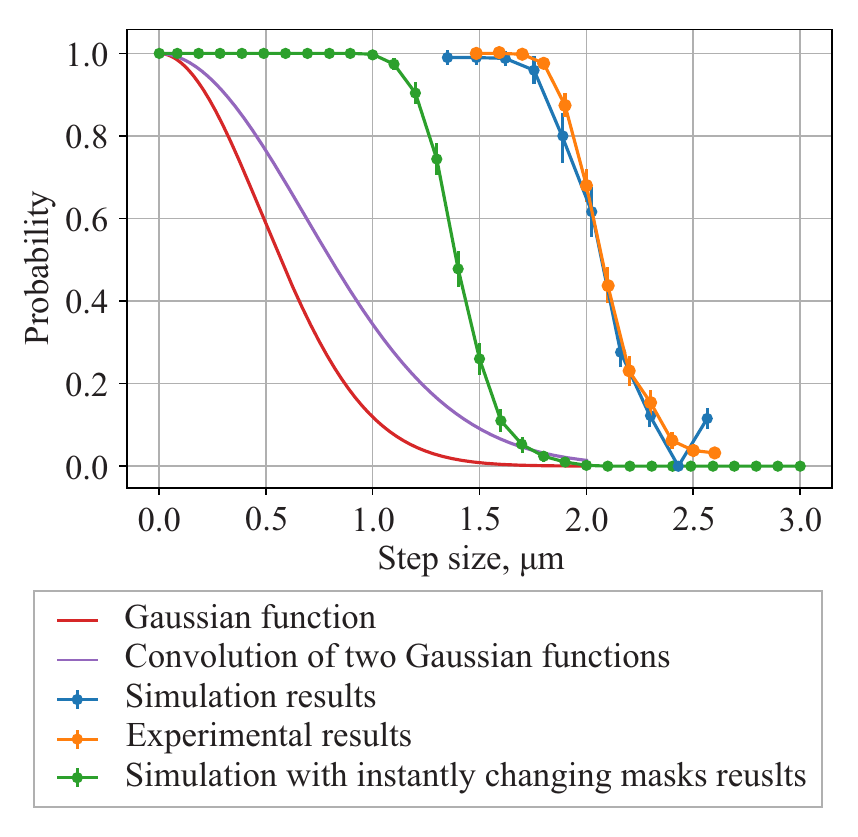} \label{fig:prob_vs_stepsize} }
	\caption{\subref{fig:lifetime_sequence}  The experimental sequence used to determine the atom lifetime in a dipole trap and survival probabilities at the end of motion. \subref{fig:prob_vs_stepsize} The results of permissible step size estimation experiment and simulations showing the dependence of the survival probability on the step size.} \label{fig:lifetime}
\end{figure}

Liquid crystal spatial light modulators typically are slow devices with frame rates ranging from tens of Hz to few kHz. Therefore, the tweezer motion has to be necessarily discretised into large steps. Faster motion requires fewer steps, therefore it is important to experimentally estimate the maximal permissible step size for atom movement. For that purpose we generated few sets of dynamic phase holograms. The dipole trap beam being imposed to them formed a triangle shaped intensity distribution in the focal plane, as shown in Figure~\ref{fig:triangle_step}. One of the dipole traps was moving back and forth in steps with a variable step size. The experimental sequence is graphically presented in Figure~\ref{fig:lifetime}\subref{fig:lifetime_sequence} and could be described as follows:

\begin{enumerate}
	\item Formation of a cold atomic cloud. Cooling, repump and dipole trap lasers are turned on.
	\item Fluorescence signal is accumulated for 100~ms, the observed number of photocounts exceeding an appropriately chosen trigger level testifies single atom trapping in the specified dipole trap and starts the following experimental sequence.
	\item Following the trigger event a dynamic phase hologram is displayed on the SLM. The cooling and repump lasers are turned off during the playback.
	\item When the playback of the hologram sequence is over and the dipole trap is returned to it's initial position, the repump and cooling laser beams are turned back on and the presence of the atom in the trap is checked by observing the fluorescence signal.
\end{enumerate}

The probability of loosing the atom form the trap is determined in a sequence of 250 repeated experiments. The experimental results are presented in Figure~\ref{fig:lifetime}\subref{fig:prob_vs_stepsize} by blue dots and line. One naturally expects this probability distribution to somehow follow the convolution of two Gaussian functions with the waists corresponding to the trapping beam waist (a violet line in Figure~\ref{fig:prob_vs_stepsize}) \cite{kim2019gerchberg}. However, the experimentally measured probability turned out to be much wider than the convolution of the beams. It means that much larger step sizes can be used than one would naively expect and the atom loss probability may still be vanishingly small.

\begin{figure}
\centering \subfigure[]{
\includegraphics[width=\linewidth]{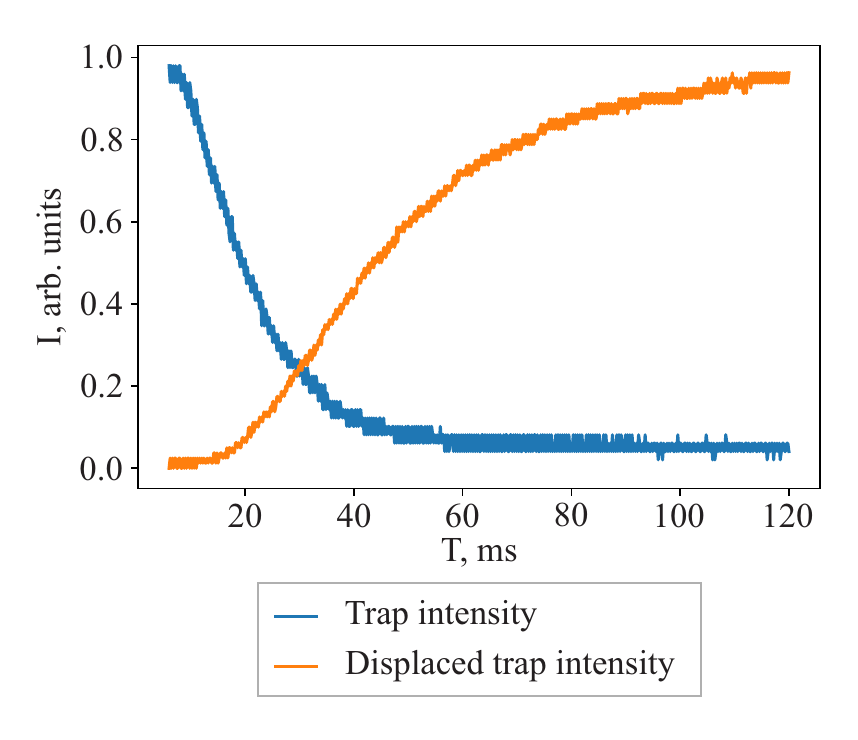}\label{fig:timing}}  
\subfigure[]{
\includegraphics[width=\linewidth]{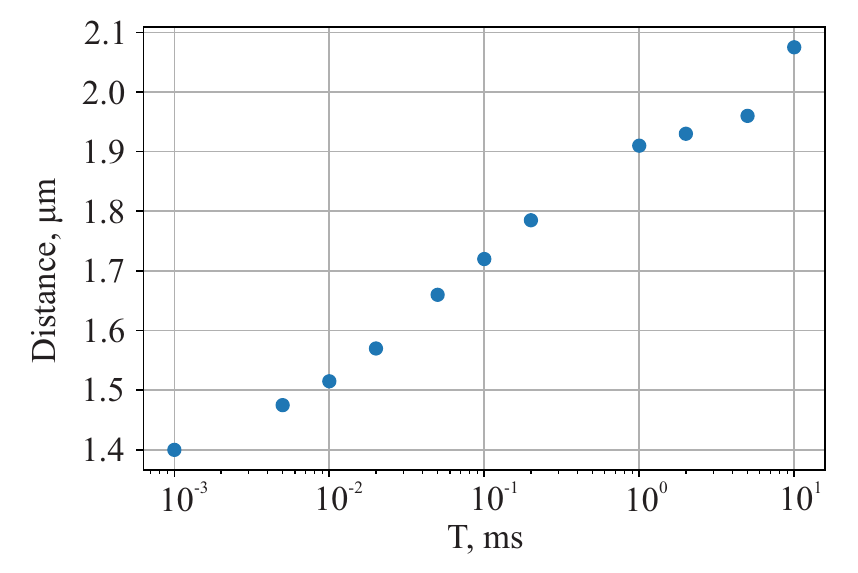} \label{fig:halftime} }
\caption{\subref{fig:timing} Experimentally measured dynamics of the trapping potential change when the two subsequent holograms corresponding to a single motion step are switched. The intensity of a dipole trap at its initial position (blue curve) and the intensity of a displaced dipole trap (orange curve). \subref{fig:halftime} Simulated dependence of the step size corresponding to the loss probability of 0.5 on the holograms switching time $T$. 
} \label{fig:sim}
\end{figure}

To explain the observed effect, we have to take into account the finite switching time of the SLM and thermal motion of the trapped atom. We experimentally characterized the dynamics of the trapping potential induced by two successive holograms in the sequence. The CCD camera in the detection part of the setup (Figure~\ref{fig:setup}) was replaced by a pair of fast photodiodes. One of them detected the signal from the trap in its initial position, while the second one measured the signal from the displaced trap. Results are presented in Figure~\ref{fig:sim}\subref{fig:timing}. It could be clearly seen that it takes around 120 ms for the single step in the sequence and the intensity of the displaced trap increases simultaneously with the intensity decay of the initial trap. Since the switching time is much larger, then the oscillation period for a trapped atom, one has to model the dynamic evolution in this adiabatically changing potential. For simplicity we used a model of linearly changing trap intensities, assuming the time dependent potential of the following form:

\begin{equation}
	\label{eq:time_dependent_potential}
	U_{tot}(t) = \left(1-\frac{t}{T}\right)U(x,y,z) +  \frac{t}{T}U(x+\delta x, y, z),
\end{equation}

\noindent where $T$ is the total switching time and $\delta x$ is the step size. Parameters of the potential were taken to correspond to a Gaussian trap with the waist od $w_0$ = 0.97 $\mu$m and longitudinal and transverse frequencies of $w_t$ = 20.8 kHz, $w_l$ = 105.2 kHz, respectively. For a trap wavelength of $\lambda$ = 852 nm the trap's depth was estimated to be $U_0$ = 1.08 mK. Shape of the trapping potential was given by the following equation:

\begin{equation}
	U(x, y, z) = \frac{U_0}{1 + \frac{z}{z_R}}\exp{\frac{-2(x^2 + y^2)}{w(z)^2}},
\end{equation}

\noindent where $w(z) = w_0\sqrt{1 + \frac{z^2}{z_R^2}}$ and $z_R = \frac{\pi w_0^2}{\lambda}$ is the Rayleight length. We performed a series of Monte-Carlo simulations of the atomic motion in a potential (\ref{eq:time_dependent_potential}). Samples were generated with coordinates and velocities taken from the normal distributions with the corresponding variances
\begin{equation}
	\sigma_{xy} = \sqrt{\frac{kT}{mw_t^2}},\quad \sigma_{z} = \sqrt{\frac{kT}{mw_l^2}}, \quad \sigma_{v} = \sqrt{\frac{kT}{m}}.
\end{equation}
The effective temperature of atoms at the beginning of the experiment was taken to be $T$ = 15 $\mu$K, as experimentally measured by a conventional release and recapture method. 

Difference between the timescales of atom oscillation frequency and SLM refresh rate of 4 orders of magnitude leads to high computational costs.
To decrease them we reduced period of masks rotation in our simulations. To make sure that such a  reduction does not affect the results of our simulation we calculated the distance where the survival probability drops below the 0.5 level as a function of $T$. This dependence is shown in Figure~\ref{fig:sim}\subref{fig:halftime}, it shows a slow logarithmic growth.

For our simulations we chose $T$ = 10~ms. Results of the simulation are shown by orange dots and line in Figure~\ref{fig:lifetime}\subref{fig:prob_vs_stepsize} and fit experimental data quite well. For comparison we show the results for an instantaneously switching potential ($T\rightarrow 0$) as a green dots and line. It is clear, that adiabatic switching results in much larger possible steps. The trade-off between the possible step size and the switching time is further illustrated in Figure~\ref{fig:halftime}. Under our experimental conditions the probability of atom survival after a single step is high enough for the step size less than 1.6~$\mu$m.

\begin{figure}
    \centering
    \includegraphics[width=\linewidth]{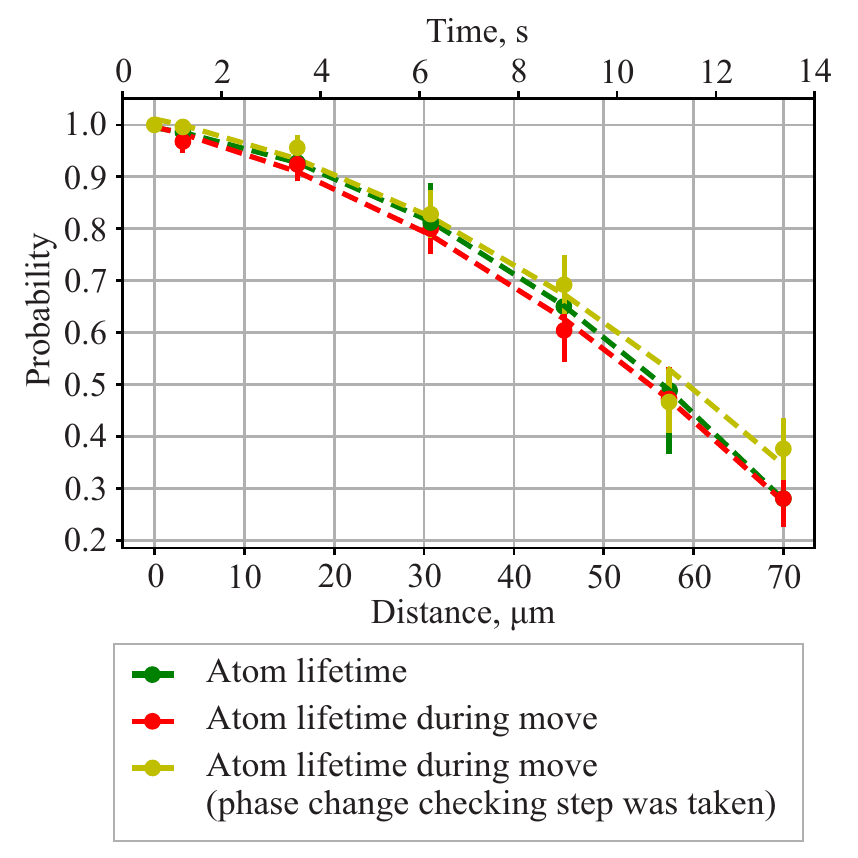}
    \caption{Atom lifetime in the static and moving traps. Atom loss probability in the static dipole trap (green dashed line and dots) and in the dynamic traps with flicker suppression by restricted phase change (yellow dashed line and dots) and without it (red dashed line and dots).}
    \label{fig:Prob_vs_scantime}
\end{figure}

For practical applications, like formation of uniformly filled atomic arrays, a dynamical tweezer should be able to transport an atom for long distances of at least several lattice periods. When this process is realized with an SLM it necessarily involves sequences of many discrete steps. We have experimentally studied the motion process and determined the loss probability for long sequences of steps. Triangular intensity distribution and straight line trajectories shown in Figure~\ref{fig:triangle_step} were used. The experimental sequence is similar to the one for a single step experiment (Figure \ref{fig:lifetime}\subref{fig:lifetime_sequence}). Sequences of holograms with flicker damping and without it were tested with the step size chosen to be 1~$\mu$m. Experimental results for the survival probabilities are shown in Figure~\ref{fig:Prob_vs_scantime}. For comparison we measured the atom lifetime in a static trap (shown by green line). Clearly the process of atom movement in the dynamical tweezer does not induce any additional loss and the probability of loss is completely determined by the atom lifetime in the dipole trap. We have also observed essentially no improvement for flicker suppressed sequences, which is probably expected, since flicker frequencies are much lower then characteristic frequencies of atomic motion in the trap. 

One may note a non-exponential character of the dependence which is a signature that we are not limited by the vacuum lifetime due to collisions with a background gas. Additional loss mechanisms are most probably related to intensity noise induced heating in the dipole trap, and may be significantly suppressed for example by introducing periodic cooling steps in the motion sequence.

We have studied the heating dynamics in both static and dynamic traps in more details by direct release-and-recapture measurements. The experimental sequence used is presented in Figure~\ref{fig:temperature_cur}\subref{fig:temperature_sequence}. It consists of the following steps:

\begin{figure}
	\centering \subfigure[]{
		\includegraphics[width=\linewidth]{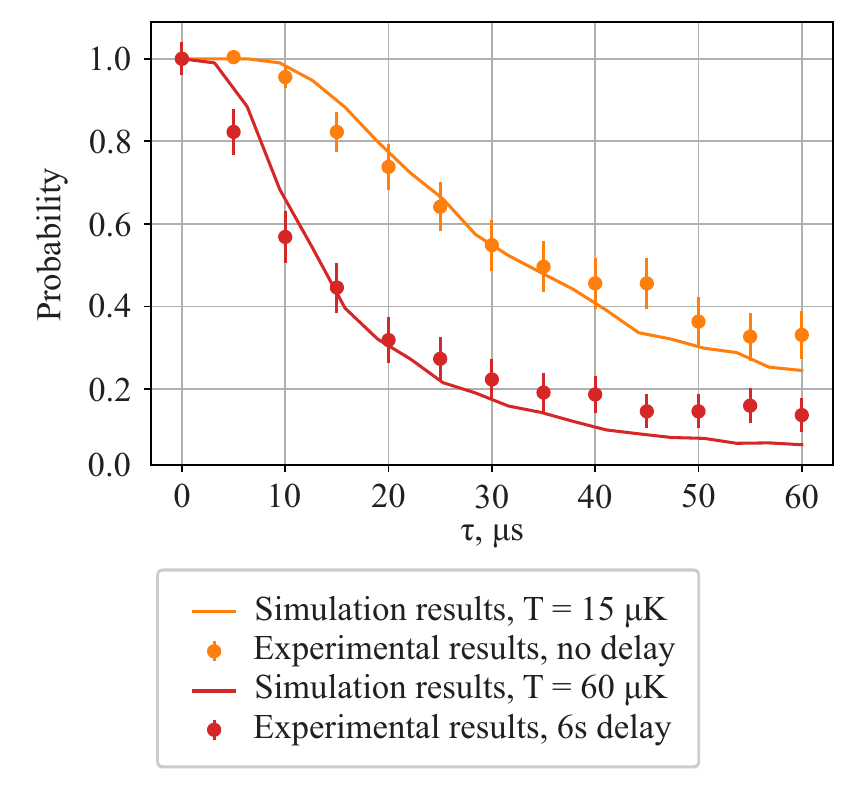} \label{fig:temperature_sequence} }  
	\subfigure[]{
		\includegraphics[width=\linewidth]{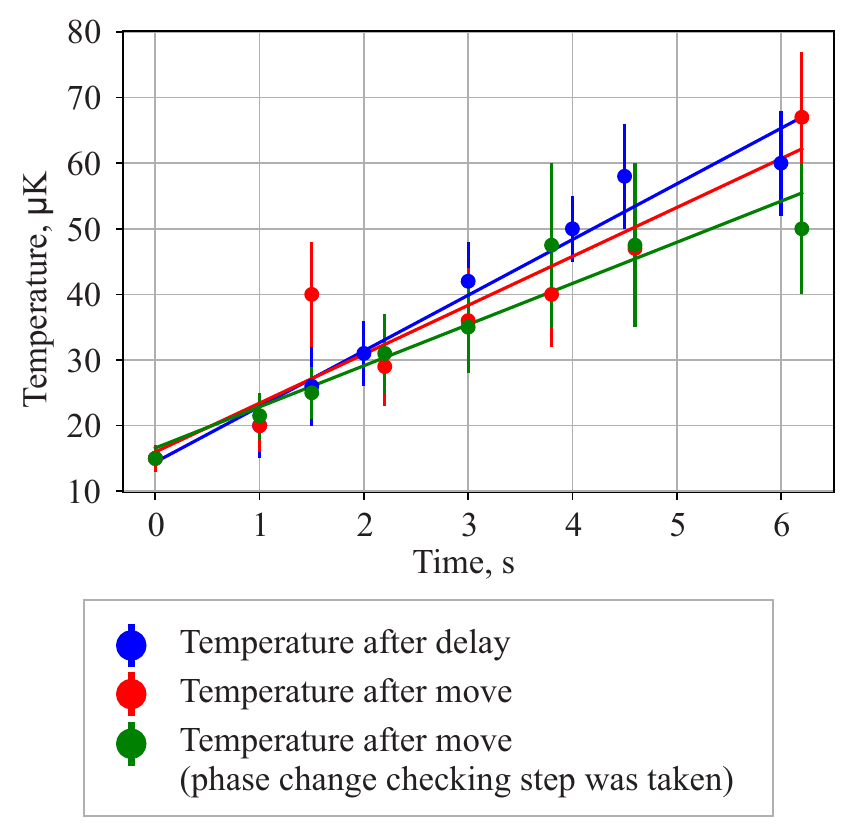} \label{fig:temperature} }
	\caption{\subref{fig:temperature_sequence} Experimental estimation of the effective temperature of an atom. Orange dots -- retention probability measured for zero delay, orange line -- Monte-Carlo simulation results for T = 15 $\mu K$, red dots -- retention probability measured for zero a 6s delay, red line -- Monte-Carlo simulation for T= 60 $\mu K$. \subref{fig:temperature} Experimentally measured heating rates. Atom temperature change in the static dipole trap (blue dots),  dynamic trap with (green dots) and without (red dots) flicker suppression. Solid lines are linear fits to the data.} \label{fig:temperature_cur}
\end{figure}

\begin{enumerate}
\item Formation of the cold atomic cloud. Cooling, repump and dipole trap lasers are turned on.
\item Fluorescence signal is accumulated for 100~ms, excess of some trigger level testifies a successful atom trapping and triggers further experimental sequence.
\item Dynamic phase hologram is displayed on the SLM. Cooling and repump lasers are turned off during the playback.
\item When the motion is finished and the dipole trap is returned to it's initial position, the dipole trap beam is turned off for a variable period of time $\tau$.
\item Dipole, repump and cooling laser beams are turned back on and the presence of an atom in the dipole trap is checked again by the fluorescence signal.
\end{enumerate}

Firstly we measured the heating rate for an atom in a static dipole trap. We have assumed, that the energy distribution for the atom has Boltzmanian form in an approximately harmonic potential \cite{Grangier_PRA2008}
\begin{equation}
	f(E)=\frac{1}{2(kT)^3}E^2\exp\left(-\frac{E}{kT}\right)
\end{equation}
with variable effective temperature $T=T(t)$, which we estimate by fitting the release-and recapture probabilities with the results of Monte-Carlo simulations. Experimental results for the initial distribution and the distribution after 6~s delay are shown in Figure~\ref{fig:temperature_sequence} along with their fit with simulated data. Both distributions are quite well simulated assuming Boltzmanian energy distributions. The dependence of the effective temperature on the delay time is shown in Figure~\ref{fig:temperature_cur}\subref{fig:temperature} and is well approximated with a linear dependence $T(t) = T_0 + \alpha t$, with the initial temperature of $T_0=15\pm2$~$\mu$K and the heating rate $\alpha = 8.5\pm 0.3$~$\mu$K/s. 

The same dependencies were measured in the dynamic holographic traps with variable motion times. No excess heating in the dynamic traps was observed, and the heating rate was approximately constant independently of the hologram sequence. We attribute this heating to high frequency intensity noises of the dipole trap laser.

\section{Conclusion}
\label{Results}

We have experimentally studied the process of single atom transport with dynamic holographic tweezers. The transport is implemented by series of holograms corresponding to shifted positions of the trapping tweezer. A remarkable experimental finding is the fact, that the shift of the tweezer in consecutive holograms may be made significantly larger, than the trap size without any increase in the probability of atom loss from the trap. This behavior is well explained by Monte Carlo simulation taking into account the dynamics of trapping potential during the change of frames and thermal motion of the trapped atom. The possibility to perform motion in large steps is essential for liquid-crystal based SLM's with slow refresh rates. 

We have experimentally studied the influence of intensity flicker that arises in dynamic holographic masks due to discrete nature of dynamic holograms on atom temperature and atom loss during movement. It turned out that the probability of successful atom movement is completely determined by the atom lifetime in the static dipole trap and is not affected by the movement process itself. We have also experimentally confirmed that the process of movement doesn't produce any additional heating in comparison with the heating induced by intensity noise of the trapping laser.

The authors are grateful to E.V.Kovlakov for enlightening discussions. This work was supported by the Russian Scientific Foundation under Grant \# 18-72-10039.

\bibliographystyle{apsrev4-1}
\bibliography{biblio}

\end{document}